\documentclass[twocolumn,secnumarabic,amssymb, nobibnotes, aps, prd]{revtex4-1}
\usepackage{graphicx}
\usepackage{color}
\usepackage{epsfig}
\usepackage{verbatim}
\usepackage{bm}

\begin{document}

\title{Terahertz Saturable Absorption from Relativistic High-Temperature Thermodynamics in Black Phosphorus}

\author{Nidhi Adhlakha$^{1}$, Zeinab Ebrahimpour$^{1,2}$, Paola Di Pietro$^{1}$, Johannes Schmidt$^{1}$, Federica Piccirilli$^{1}$, Daniele Fausti$^{3,4}$, Angela Montanaro$^{3,4}$,  Emmanuele Cappelluti$^5$ , Stefano Lupi$^6$ and Andrea Perucchi$^{1}$}

\affiliation{$^1$Elettra - Sincrotrone Trieste S.C.p.A, S.S. 14 km163.5 in AREA Science Park, 34012 Trieste, Italy}
\affiliation{$^2$Abdus Salam International Centre for Theoretical Physics, Strada Costiera 11, Trieste I-34151, Italy}
\affiliation{$^3$Department of Physics, Universit\`a degli Studi di Trieste, 34127, Trieste, Italy}
\affiliation{$^4$Chair of Solid State Physics, Department of Physics, University of Erlangen-N\"urnberg, 91058 Erlangen, Germany}
\affiliation{$^5$Istituto di Struttura della Materia, CNR (ISM-CNR), 34149 Trieste, Italy}
\affiliation{$^6$CNR-IOM  and Dipartimento di Fisica, Universit\`a di Roma Sapienza,  P.le Aldo Moro 2, I-00185 Roma, Italy}

\date{\today}


\begin{abstract}

Thanks to its tunable infrared band-gap and to its anisotropic conduction properties, black phosphorus represents a very unique 2D material, whose potential in the engineering of new devices still needs to be fully explored. We investigate here the nonlinear terahertz (THz) electrodynamics  of black phosphorus along the more conducting armchair direction. Similarly to the case of other 2D systems like graphene and topological insulators, the THz saturable absorption properties of black phosphorus can be understood within a thermodynamic model by assuming a fast thermalization of the electron bath. While black phosphorus does not display the presence of massless fermions at ambient pressure and temperature, our analysis shows that its anomalous THz nonlinear properties can be accounted for by a relativistic massive Dirac dispersion, provided the Fermi temperature is low enough. An optimal tuning of the Fermi level therefore represents a strategy to engineer strong THz nonlinear response in other massive Dirac materials as in transition metal dichalchogenides or high-temperature superconductors.

\end{abstract}

\maketitle

\section{Introduction}
\label{introduction}

Among the family of 2D materials black phosphorus (BP) stands out for its highly peculiar properties. At a fundamental level, BP is a very fascinating material due to the occurrence of a pressure-induced topological Lifshitz transition, which turns the material from semiconducting to metal \cite{xiang15,dipietro18}. It is found indeed that at relatively low-pressures ($\sim1.5$ GPa) a non-avoided band crossing gives rise to a plasma of Dirac massless charge carriers. 

On the other hand, BP is also extremely appealing for opto-electronic applications \cite{montanaro22}, since it couples a significantly high mobility (reaching up to 1000 cm$^2$/V.s), with the presence of an infrared and tunable ($0.3-2$ eV) band-gap \cite{ling15}. 
Its unique anisotropic in-plane transport, may be further exploited to design devices with completely new functionalities \cite{low14,fei14}. BP is also a hyperbolic photonic material in the THz range \cite{biswas21}, a property which can be exploited for a large variety of applications as for instance hyperlensing or sub-diffraction light confinement. Nonlinear effects are crucial in many opto-electronic applications, as for instance for ultra-fast signal processing \cite{li14a} or optical sensing \cite{peters22}.

Remarkably, miniaturization and electromagnetic confinement will induce the presence of strong electric fields which may affect BP in a nonlinear way, thus making its optical properties dependent on the characteristics of the applied THz beam.
In the dc limit BP is also known to exhibit nonlinearities, varying from current saturation to impact ionization \cite{li14,wang14,das14,ahmed18}.

The physical mechanisms underlying these nonlinearities is still debated, and may be intimately connected to the low frequency limit of the conductivity in the THz regime. The saturation properties in the THz absorption displayed by 2D materials like graphene or topological insulators were recently investigated both theoretically and experimentally by making use of high peak power THz sources \cite{hafez20,giorgianni16}. One may thus wonder whether BP, being at the verge of a Lifshitz transition which would eventually give rise to Dirac electrons, may share the same physics. 

To answer this question we provide here the first characterisation of the nonlinear optical properties of bulk BP as a function of the incoming THz field, by exploiting the high-power TeraFERMI \cite{perucchi13,dipietro17} source. This characterisation serves as a benchmark for theoretical models, which aim to identify the fundamental ingredients responsible for strong THz nonlinear behaviour in general.

\begin{figure*}
\begin{center}
\leavevmode
\includegraphics [width=15cm]{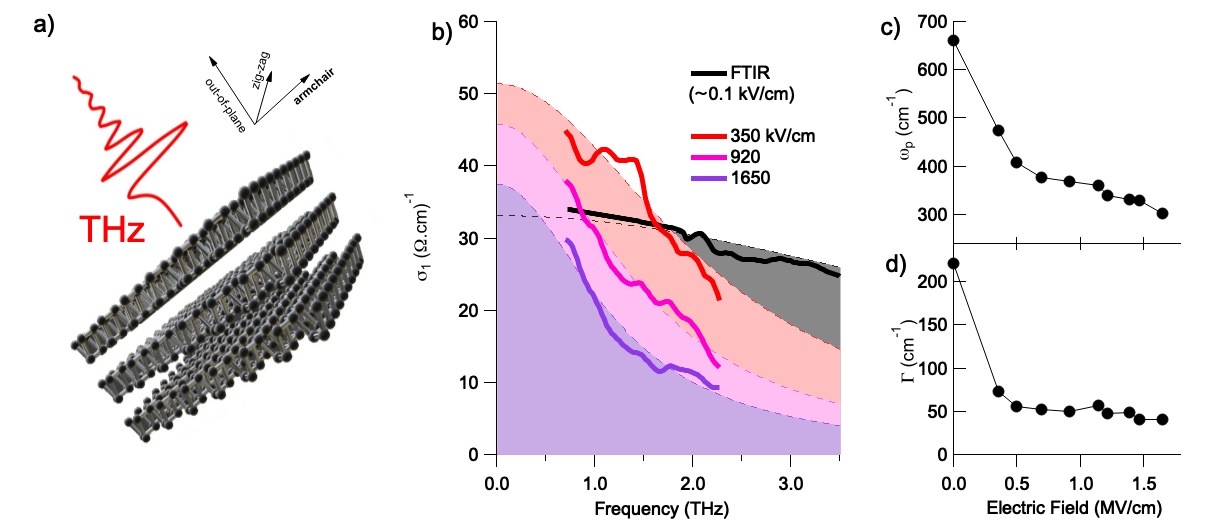}  
\end{center}
\caption{a) Black Phosphorus structure. b) Real part of the optical conductivity of black phosphorus measured along the  {\it armchair} polarization directions 
at selected THz electric field strengths. The measurement at low THz field ($\sim 0.1$ kV/cm) was performed with a Fourier Transform Infrared (FTIR) spectrometer, while the measurements at selected field strengths were performed by employing the single-cycle pulses from the TeraFERMI facility. The overall behavior of $\sigma_1(\nu)$ can be described (dashed lines) by the Drude model. The optical conductivity decreases when the THz field increases. THz field dependent plasma frequency $\omega_p$ (c) and scattering rate $\Gamma$ (d), as extracted from the Drude fitting.}
\label{Fig1}
\end{figure*}

\section{Results and Discussion}

The main result from our investigation is summarized in Fig. \ref{Fig1}, where we compare the real part of the linear (low electric THz field) optical conductivity $\sigma_1(\nu)$, as extracted from FTIR reflectivity measurements at the SISSI beamline \cite{lupi07}, with the conductivity for three selected high THz field intensities. These measurements are performed with THz light polarized along the more conducting  {\it armchair} direction, while results from the {\it zig-zag} polarization direction are briefly discussed in Appendix \ref{Experimental}. The THz optical conductivity for all fields presents a weakly metallic behavior due to the presence of dopant defects \cite{dipietro18}. We notice, from Fig. \ref{Fig1}b  that $\sigma_1(\nu)$ is not constant for all incoming THz fields, but progressively decreases as long as the THz electric field value increases. 

The optical conductivity can be fitted with one single Drude term, thus allowing to estimate the plasma frequency ($\omega_p$) and scattering rate ($\Gamma$),  as reported in Fig. \ref{Fig1}c-d. An analysis of the Drude parameters reveals that the observed decrease of $\sigma_1(\nu)$ can not be simply attributed to an enhanced scattering rate associated to heating effects. On the contrary, the scattering rate decreases for higher fields, an effect previously observed in graphene, and attributed to the dominance of long-range scattering on Coulomb impurities \cite{mics15}. This scenario is even more likely in BP where significant concentrations of ionized point defects are believed to act as very efficient charge-carrier scattering agents \cite{liu17}.

The decrease of the optical conductivity at high fields is therefore driven by the reduction in the plasma frequency $\omega_p=\sqrt{4\pi Ne^2/m^*}$ where $N$ is the carrier density, and the effective mass is defined as $m^*=\frac{\hbar^2}{\mathrm {d}^2\varepsilon/\mathrm{d}k^2}$. The plasma frequency measured along the {\it armchair} direction varies between 660 cm$^{-1}$ at 0.1 kV/cm to 470 and 300 cm$^{-1}$ at 350 and 1650 kV/cm, respectively. This shows that strong nonlinear effects are already at play between $\sim0.1$ and 350 kV/cm.

The effective mass $m^*$ used for the calculation of $\omega_p$ is a constant only as long as the charge carriers are restricted to a small portion of the Brillouin Zone (BZ) at the bottom of the conduction band, where the parabolic approximation holds. This is however no longer true for high accelerating fields which can drive charge carriers in regions of the BZ where the band dispersion relation is no longer parabolic. This is indeed the case for InSb \cite{yu17,houver19} or Bi \cite{minami15}, where the breakdown of the effective mass approximation explains their saturable absorption properties in the presence of strong THz fields.

This scenario can be mimicked with a Finite-Difference Time-Domain FDTD approach, which incorporates the nonlinear effects by making use of a wavevector dependent effective mass $m^*=m^*(\bm{k})$, as in Ref. \cite{yu17}. As detailed in Appendix \ref{FDTD}, the calculation qualitatively reproduces the enhancement of the transmitted THz pulses for increasing incoming fields, but overestimates the transmission increase, especially for fields higher than 1 MV/cm. 

Here one should keep in mind that FDTD is a purely one-electron model, which incorporates scattering only as a friction parameter in the quasi-classical electron's equation of motion. However, for sufficiently high fields, both electron-electron and electron-phonon scattering can have dramatic effects on the THz nonlinear properties. As a consequence, in most realistic cases, temperature effects need to be incorporated in some form. 

\begin{figure*}
\begin{center}
\leavevmode
\includegraphics [width=18cm]{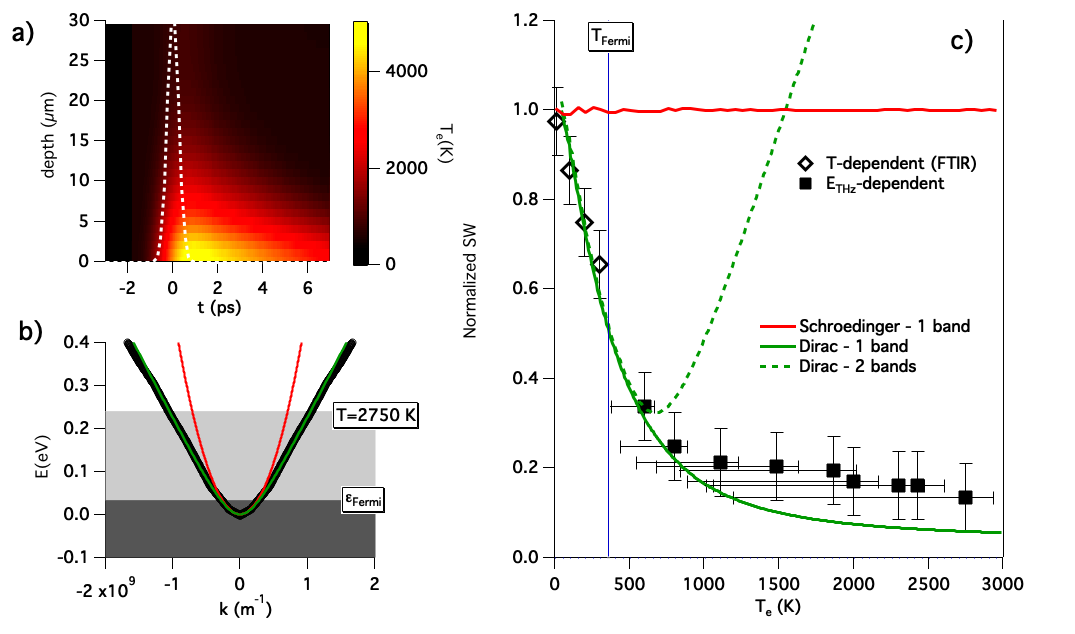}  
\end{center}
\caption{ (a) False color representation of the electron temperature as a function of time (horizontal) and penetration inside the sample (vertical), calculated for the {\it armchair} polarization at the maximum applied THz electric field (1650 kV/cm). The dashed white line is a gaussian with a FWHM of 590 fs representing the time profile of the THz source used in the simulation. (b) Energy vs momentum dispersion of BP from Reference \cite{ehlen16} (black), and for three model dispersion relations: Schr\"odinger (red) and Dirac massive (green). 
(c)  Normalized SW (markers) as a function of the $T_e$, for $T$-dependent reflectivity (diamonds), and THz field-dependent data (squares). To take into account possible scattering events occurring in different directions of the k space we set as a lower bound of $T_e$, the value obtained from a simulation for an effective "isotropic" electron heat capacity, calculated as a geometrical average over the {\it armchair} and {\it zig-zag} directions. The continuous lines correspond to the calculated $SW(T)$ for the two model dispersion relations as in (b). The dashed green light corresponds to the Dirac model allowing for electron-hole pairs formation (see text).}
\label{T_mu_bands}
\end{figure*}

\section{Thermodynamic Model}

The most successful approach used to describe THz nonlinear properties of quantum materials, is the thermodynamic model introduced by Mics {\it et al.} \cite{mics15} to explain the THz induced transparency in graphene. Besides reproducing the optical conductivity of graphene at high fields, the thermodynamic model demonstrated its validity also in modeling THz harmonics generation  \cite{hafez18}, and plasmon resonance softening both in graphene \cite{jadidi19} and topological insulators \cite{dipietro20} in ribbon array structures. The thermodynamic model assumes that the electrons which are absorbing energy from the THz field quickly exchange energy through electron-electron collinear scattering events \cite{hafez20}, while the lattice temperature remains the same. As a consequence of the ultrafast electron heating, the chemical potential readjusts to the increased electron temperature ($T_e$) thereby lowering its Drude weight.

To check whether the thermodynamic model can be applied to BP as well, we first evaluate the heating induced by the THz pulses. To this aim, we employ a two-temperature model,
\begin{equation}
c_e\rho^{BP}\frac{\partial T_e}{\partial t}=G (T_l-T_e)+S(t),
\end{equation}
\begin{equation}
c_l\rho^{BP}\frac{\partial T_l}{\partial t}=G (T_e-T_l), 
\end{equation}
where $S(t)$ corresponds to the impulsive heat provided by the THz pulse, $G$ is the electron-phonon coupling constant, and $c_e(T)$ and $c_l(T)$ are the electronic and lattice components of the heat capacity. The system of coupled differential equations is solved with the help of the NTMpy package \cite{alber20}. The details of the parameters are provided in Appendix \ref{2T}. We note however that the electron heat capacity is evaluated through the textbook formula $c_e(T)=\frac{1}{3}\pi^2D(\varepsilon_F)k_B^2T$, where $D(\varepsilon_F)=\frac{3}{2} N/\varepsilon_F$. The Fermi energy $\varepsilon_F=32$ meV is estimated by making use of low temperature FTIR data, as described in Appendix \ref{FTIRdata}. 

The results of the two-temperature model are reported in Fig. \ref{T_mu_bands}a, showing the distribution of the electronic 
temperature in time and space, along the whole thickness of our BP sample, for the highest THz field (1.65 MV/cm). In order to evaluate an effective temperature for the whole sample we perform an averaging along the entire sample thickness, weighted by the THz pulse penetration depth ($\sim 6$ $\mu$m).

The two-temperature model thus allows us to plot the evolution of the Drude spectral weight ($SW=\omega_p^2$) as a function of $T_e$, ranging from 600 to 2750 K. This can be compared with the $T$-dependent data from FTIR infrared reflectivity (see Appendix \ref{FTIRdata}), covering on the other hand $T$ values from 10 to 300 K.
The results are reported in (Fig. \ref{T_mu_bands}c), where the Drude weight is plotted either as a function of $T$  for FTIR, or as a function of $T_e$ for the THz field-dependent data.

Overall, the spectral weight  decreases monotonically as a function of temperature, with a striking drop observed between the FTIR and the field-dependent experimental data. The collapse of the spectral weight is remarkable, since at the highest electric field value, corresponding to $T_e \sim 2750$ K, the $SW(T)$ has dropped below 20\% of its low-temperature value. Interestingly, the Fermi temperature $T_F=370$ K coincides with the region where the $SW(T)$ decrease is more pronounced. This indicates that $T_F$ sets the temperature scale for the dramatic decrease of the spectral weight.

We now want to theoretically estimate the expected $T$-dependence of the $SW$. To this aim, we first need to establish the $T$-dependence of the chemical potential $\mu(T)$. Due to the high values of $T_e$ considered here, ($T_e \gtrsim 7 T_F$),  the usual Sommerfeld expansion cannot be employed. $\mu(T)$ is therefore evaluated through a numerical procedure \cite{jadidi19} calculating the shift necessary to conserve the total number of charge carriers, according to the Fermi-Dirac distribution (see Appendix \ref{SW_T}):
\begin{equation}
N(T)=N(T=0)=\int^{+\infty}_{-\infty}{g(\varepsilon)f(\varepsilon,\mu(T),T)d\varepsilon},
\end{equation}
where $f(\varepsilon,\mu(T),T)=(1+exp[\frac{\varepsilon-\mu(T)}{k_BT}])^{-1}$, and $g(\varepsilon)$ is the density of states. Once the temperature-dependence of the chemical potential has been established we can calculate $SW(T)$ as
\begin{equation}
SW(T)\propto\int^{+\infty}_{-\infty}v(\varepsilon)^2 g(\varepsilon)\frac{\partial f(\varepsilon,\mu(T),T)}{\partial \varepsilon}d\varepsilon,
\label{SW}
\end{equation}
where $v(\varepsilon)$ is the electron velocity, defined by 
$v(\varepsilon)=\frac{\partial \varepsilon(k)}{\partial k}|_{k_0:\varepsilon(k_0)=\varepsilon}$.

It remains to be seen what is the form of the density of states $g(\varepsilon)$ which better reproduces our experimental results.
As shown in Fig. \ref{T_mu_bands}b, a parabolic Schr\"odinger-like dispersion with $m^*=0.08 m_e$ fits the BP band-structure over a rather limited wavevector range, roughly corresponding to the energy scale set by $\varepsilon_F$. At higher energies the parabolic dispersion strongly deviates from the real band structure of BP. On the other hand, making use of a massive Dirac dispersion in the form $\varepsilon=\sqrt{m^{*2}c^4+p^2c^2}$, allows a better matching to  the actual BP band structure over an extended energy range, from 0 to 0.4 eV ($k\sim0 - 1.6$ nm$^{-1}$). 

The choice of the proper band dispersion has a profound impact on the functional dependence of $g(\varepsilon)$ and $v(\varepsilon)$, and therefore on the evaluation of $SW(T)$, as demonstrated in Fig. \ref{T_mu_bands}c). The most striking feature is that the Schr\"odinger-like parabolic band model predicts a T-independent Drude spectral weight. This is a consequence of the so-called f-sum rule 
\begin{equation}
SW(T)=\omega_p(T)^2=8\int^{\infty}_0\sigma_1(\nu)d\nu=\frac{4\pi Ne^2}{m^*}, \label{sumrule}
\end{equation}
representing a statement on the conservation of the particle's number.  This implies that in a model considering one single (infinite) band in a purely parabolic potential, $SW(T)$ can not change. It is known however that relation (\ref{sumrule}) is no-longer true for Dirac materials \cite{gusynin07,sabio08,throckmorton18}. In the case of graphene, both the compensated semimetal properties as well as the Dirac nature of the quasi-particles are responsible for the peculiar $T$-dependence of the $SW(T)$ \cite{frenzel14}. 

For an infinite 3D Dirac dispersion, the numerical calculation shown in Fig. \ref{T_mu_bands}c predicts a dramatic drop of the Drude $SW(T)$ in good agreement with our experimental observation. 
This shows that the THz nonlinear properties of BP are ruled by the high temperature ($T>T_F$) thermodynamics of a relativistic Fermi gas \cite{sevilla17} which was previously addressed in the framework of the study of white dwarf stars, hot quark matter and gluon-quark plasma, rather than condensed matter physics.

In this regard, a very interesting question is whether the particle-antiparticle symmetry should be included in our model, by taking into account the thermodynamic equilibration with holes in the valence band. This phenomenon could take place via impact ionization \cite{ahmed18}, when charge carriers accumulate enough energy from the driving field, so that they can be promoted in the conduction band. We have therefore calculated the thermodynamic $SW(T)$ dependency by taking into account the presence of both electron and hole's dispersions. In this case, depicted by the green dashed line in Fig. \ref{T_mu_bands}c), an upturn in $SW(T)$ would be expected at $T_e\sim 600$ K, while $SW(T)$ may even exceed its low-temperature value for $T_e\gtrsim1500$ K. 

This scenario is in disagreement with our experimental findings. We believe that the reason for the discrepancy is that impact ionization is an avalanche phenomenon requiring the accumulation of collisions, so that thermalization of electron-hole pairs can not happen on the sub-ps time-scale set by the duration of the THz pulse itself. It is therefore likely that for the full impact ionization process to take place, a longer time-scale is needed probably in the 10's ps range, as previously observed for InSb \cite{hoffman08}.
Nonetheless, the onset of the impact-ionization phenomenon may provide a qualitative explanation for the small differences between the experimental $SW(T)$ and the predictions of the thermodynamic model for one single Dirac band, i.e. without electron-hole pairs formation (full green curve in Fig. \ref{T_mu_bands}c). Future time-resolved THz-pump/THz-probe experiments may provide deeper insight in the nonlinear phenomena taking part.

\section{Concluding Remarks}

We have shown that a very simple thermodynamic model within the massive Dirac band dispersion - while disregarding the microscopic details of electronic transport at high fields - can be successfully applied to quantitatively describe the THz saturable absorption properties of bulk black phosphorus.
This result is particularly interesting since the relativistic treatment is normally not needed to account for BP's properties at room temperature and pressure. 
Thanks to the unique combination of massive Dirac dispersion and low $T_F$, THz light can nevertheless be used to drive BP in a high temperature thermodynamic regime which was hitherto confined to cosmology rather than condensed matter physics.
Our findings have important consequences on the design of black phosphorus-based opto-electronic devices, as well as on the engineering of novel THz nonlinear materials.

\section*{Acknowledgments}

This work was supported by the MIUR through the PRIN program No. 2017BZPKSZ. S.L. was supported by MIUR through the PRIN programs No.:2020RPEPNH and by the PNRR MUR project PE0000023-NQSTI. E.C. acknowledges financial support from PNRR MUR project PE0000023- NQSTI. A.P. acknowledges G. Perucchi for the artwork in Fig.1.

\appendix

\section{Sample}
\label{Sample preparation}

High quality black phosphorus crystalline samples with a purity $>99.995 \%$ were purchased from HQ Graphene ({\it www.hqgraphene.com}). The sample was cleaved with Scotch tape until reaching a final thickness $d=30 \pm 10$ $\mu$m, as measured with a caliper. The sample was then quickly glued on a sample holder and mounted in the TeraFERMI set-up, where it was kept under N$_2$ purging conditions ($\leqslant 3\%$ humidity) during the whole measurement.

\section{The TeraFERMI THz source}
\label{source}
The THz source used for the THz electric field dependent measurements is the TeraFERMI superradiant THz beamline at the FERMI free-electron-laser facility \cite{perucchi13,dipietro17}. TeraFERMI employs sub-ps electron bunches in the nC range to emit THz light through the Coherent-Transition-Radiation mechanism induced by a 1 $\mu$m-thick Al membrane. THz pulses are generated with repetition rate at 50 Hz and are then guided from the source in vacuum to the TeraFERMI endstation exploiting the high peak electric fields for non-linear spectroscopy.

The spectrum of the source is acquired through EOS, %by employing a 1 mm ZnTe crystal after strongly attenuating ($\sim$10$^{-3}$) the THz pulse intensity. 
We report in Fig. \ref{source} the time profile as well as the spectral content of the pulse, as the squared amplitude of the Fourier Transform. The intensity of the THz pulses was measured by utilizing a pyroelectric detector previously calibrated with a GENTEC THZ12D powermeter. The maximum intensity measured at sample position was $I=9$ $\mu$J. The spatial profile was characterized with the help of a Pyrocam III THz camera, yielding a radius $r=305$ $\mu$m. The pulse length is approximated from the measured THz electric field time-trace with a gaussian fit of its intensity (squared electric field). The maximum electric field strength is then estimated assuming also a gaussian spatial intensity distribution with an area of $A=\pi r^2$:

\begin{equation}
E_0=\sqrt{\frac{\eta I 2ln(2)}{\pi r^2\Delta t}}=1.65 \: {\rm MV/cm},
\end{equation}
where $\eta=$ 377 $\Omega$ is the free space impedance.

\begin{figure*}
\begin{center}
\leavevmode
\includegraphics [width=15cm]{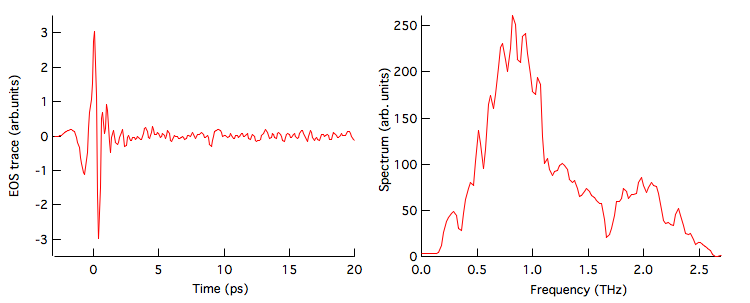}  
\end{center}
\caption{Representative THz time trace of the source (left) and corresponding intensity spectrum (right)}
\label{source}
\end{figure*}

\begin{figure*}
\begin{center}
\leavevmode
\includegraphics [width=7cm]{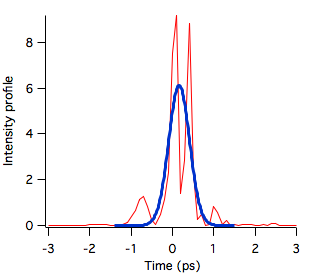}  
\end{center}
\caption{Gaussian fit of the intensity time profile. The pulse duration is estimated $\Delta t= 590$ fs FWHM.}
\label{gaussian}
\end{figure*}

\section{Experimental}
\label{Experimental}

The THz-field dependent optical conductivity data are extracted from a transmission experiment performed at different fluences of the incoming THz beam. To this aim we attenuate the THz pulses with a set of three photolithographic polarizers (from QMC Instruments and Tydex). The first and third polarizers are kept at the same orientation, while the central one is allowed to rotate (at an angle $\theta_i$) thereby attenuating the electric field according to Malus law. 

The THz light transmitted by the black phosphorus (BP) sample is detected with an electro-optic sampling (EOS) set-up, based on a 1 mm thick ZnTe crystal and a 79.9 MHz C-Fiber780 laser from MENLO, optically synchronized to the FERMI master-clock. As a reference, we measure the EOS signal (with the same ZnTe crystal) with empty sample-holder, after strongly attenuating the signal with the polarizers ($\sim 5\times10^{-2}$) to the angle $\theta_{min}$, corresponding to the lowest THz intensity in the present experiment.

The reference EOS trace is scaled with the peak value recorded in an EOS measurements performed for all attenuation angles with a GaP 100 $\mu m$-thick crystal instead of ZnTe. 
\begin{equation}
%E^{ref}(\theta_i)=E^{ref}_{ZnTe}(\theta_{min})*\frac{E^{ref,max}_{GaP}(\theta_i)}{E^{ref,max}_{GaP}(\theta_{min})}
E^{ref,\theta_i}(t)=E^{ref,\theta_{min}}_{ZnTe}(t)\cdot\frac{E^{ref,\theta_i}_{GaP}(t_{max})}{E^{ref,\theta_{min}}_{GaP}(t_{max})}
\end{equation}

The use of GaP, as a normalization for the reference spectra, avoids incurring in saturation problems of the more sensitive ZnTe crystal when the THz pulses are not attenuated enough by the polarizers, or by the sample itself. This allows evaluating the transmission as
\begin{equation}
T_i(t)=E^{BP,\theta_i}_{ZnTe}(t)/E^{ref,\theta_i}(t)
\end{equation}

The real and imaginary part of the refractive index $\tilde{n}(\nu)=n(\nu)-ik(\nu)$ are evaluated using standard formulas, as described in Ref. \cite{jepsen19}.

 The real part of the optical conductivity is finally evaluated through

\begin{equation}
%\sigma_1(\omega)=\frac{1}{4\pi}2n(\omega)k(\omega)
\sigma_1(\nu)=\frac{1}{4\pi}2n(\nu)k(\nu)
\end{equation}

The optical conductivities extracted as discussed above, were fitted between $\nu=0.5$ and 2.5  THz, by employing a simple Drude model:
\begin{equation}
\sigma_1(\nu)=\frac{1}{4\pi}\frac{\omega_p^2}{\Gamma^2-(2\pi\nu)^2}
\label{drude}
\end{equation}

\begin{figure*}
\begin{center}
\leavevmode
\includegraphics [width=15cm]{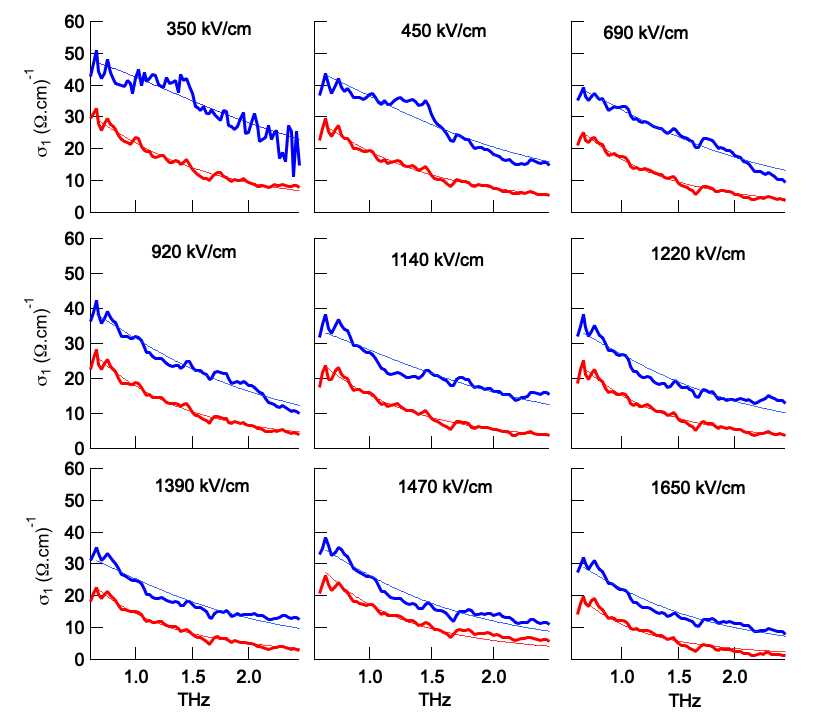}  
\end{center}
\caption{Real part of the optical conductivity, at all measured electric field values along the {\it armchair} (blue) and {\it zig-zag} (red) polarization directions. The lighter lines correspond to the Drude fittings. }
\label{sigma}
\end{figure*}

Figure \ref{sigma} reports on the optical conductivity measured along both {\it armchair} (blue) and {\it zig-zag} (red) polarization directions. In the case of {\it zig-zag} polarization, a less pronounced saturable absorption behavior was found if compared with the case of {\it armchair} direction. This result can be qualitatively understood by considering the reduced anharmonicity of the {\it zig-zag} band dispersion with respect to the Dirac-like {\it armchair} polarization, as discussed in the main text.

\section{T-dependent FTIR Reflectivity data}
\label{FTIRdata}

Temperature-dependent Reflectivity measurements were performed on a bulk sample from the same batch as the one used for the THz-field dependent measurements. A bulk freshly cleaved sample was mounted on a Helitran LT-s He-flux cryostat equipped with different optical windows (polyethylene and KRS5 for the far- and mid-infrared ranges respectively). The measurements were performed at the SISSI infrared beamline \cite{lupi07} at nearly normal incidence by employing a Bruker Vertex 70v FTIR spectrometer, equipped with suitable beamsplitters (Si, KBr) and detectors (Si-bolometer, MCT photodetector). A linear polarizer is inserted in the optical path to select the response from the {\it armchair} direction. Such orientation is selected by maximizing the reflectivity in the far-infrared range. As a reference for the reflectivity measurement we employ the gold-overcoating technique \cite{homes93}.

\begin{figure*}
\begin{center}
\leavevmode
\includegraphics [width=7cm]{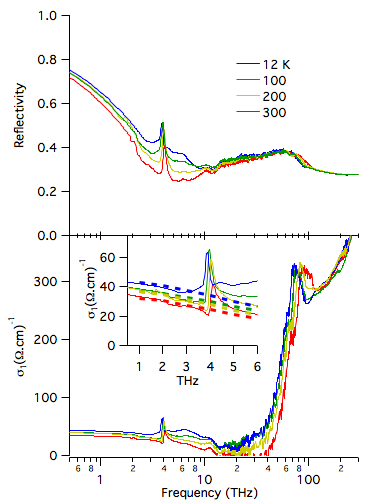}  
\end{center}
\caption{Temperature-dependent infrared reflectivity (upper panel), and real part of the optical conductivity (lower panel) as extracted from Kramers-Kronig Transformations.}
\label{FTIR}
\end{figure*}

The optical reflectivity data are extrapolated at low and high frequencies using standard procedures, in order to extract the optical conductivity through Kramers-Kronig transformations \cite{Dressel}. The THz optical conductivity is fitted with the help of the Drude model, as in equation (\ref{drude}), thus allowing to establish the temperature-dependence of the plasma frequency $\omega_p(T)$ and scattering rate $\gamma(T)$.

From the value of the plasma frequency $\omega_p(T)$, and by using effective mass  from literature $m^*/m_e=$ 0.08 for the {\it armchair}  direction \cite{liu16,narita83}, it is possible to calculate the charge density:
\begin{equation}
N=\frac{\omega_p^2m^*}{4\pi e^2}=5.8\times 10^{17} \: {\rm cm}^{-3}.
\end{equation}
The Fermi energy can be then estimated through:
\begin{equation}
\varepsilon_F=\frac{\hbar^2}{2m^*}(3\pi^2N)^{2/3}=31.7 \: {\rm meV}.
\end{equation}

The density of states at the Fermi level is finally given by:
\begin{equation}
D(\varepsilon_F)=\frac{3}{2}\frac{N}{\varepsilon_F}= 1.71\times10^{44} \: {\rm J}^{-1}\: {\rm m}^{-3}.
\end{equation}

\section{Nonlinear FDTD model}
\label{FDTD}

A one-dimensional finite-difference time-domain (FDTD) method based on the Yee algorithm together with the model of the ballistic motion of conduction electrons, as presented in \cite{yu17}, can be used to describe the intense THz pulse-induced transparency observed in the bulk black phosphorus (BP) in semi-quantitative agreement with experiment. According to this model, under the intense THz field, the electrons are accelerated to the highly nonparabolic regions of the conduction band energy of BP in the first Brillouin zone. The model does not take into account the interband tunneling, impact ionization, or any scattering mechanisms other than those in the Drude model. In this theory, the electric displacement ($D$) in black phosphorus (BP) due to the propagation of the THz beam is given by
\begin{equation}
D=\epsilon_0\epsilon_\infty E_{THz}+P_{NL} \label{s1}
\end{equation}

 where $E_{THz}$ is the electric field of the THz beam, $\epsilon_0$ is the permittivity of free space and $\epsilon_{\infty}=1+\chi_0$ is the background dielectric constant, with $\chi_0$ being the background high-frequency dielectric susceptibility. $P_{NL}$ is the nonlinear (NL) polarization arising from the conduction electron responses to the extreme THz beam. The temporal evolution of the polarization, $P_{NL}$, is defined as 
 \begin{equation}
 \frac{dP_{NL}}{dt}=-Nev_g (k)	\label{s2}
 \end{equation}
 with $v_g (k)$ being the group velocity of an electron wave packet of wave vector $k$. Using a semi-classical description, $v_g (k)$ is given by

\begin{equation}
v_g(k)=\frac{1}{\hbar} \frac {\partial\varepsilon(k)}{\partial k} \label{s3}
\end{equation}
where $\varepsilon(k)$ is the conduction band energy-momentum dependence. 
The temporal evolution of wave vector $k$ is governed by the equation of motion of the electron in the response to the THz electric field:
\begin{equation}
\frac{dk}{dt}+\Gamma k=\frac{e}{\hbar} E(z,t)	\label{s4}
\end{equation}
where $\Gamma$ is the electron scattering rate in the linear regime. In equation (\ref{s2}), $n$ is the carrier density which is obtained from the plasma frequency relation with the electron density $n$ and electron effective mass $m^*$, i.e., $\omega_p^2  =Ne^2/(\epsilon_0 \epsilon_\infty m^*)$.
The FDTD method was used to solve the time-dependent Maxwell equations for the propagating THz electromagnetic fields ($E$, $D$, and $H$) inside the BP slab of thickness $d= 30$ $\mu$m. First, the $H$ and $D$ are calculated by solving the Maxwell curl equations for a propagating THz field using the Yee central difference FDTD algorithm in the time step n+1 using the values at the earlier time steps. Then, the electric field $E$ for the time step n+1 at each point on the spatial grids s of the BP slab position is given by

\begin{equation}
E_s^{n+1}=D_s^{n+1}P_{NL}^{n+1}/\epsilon_\infty\epsilon_0 \label{s5}
\end{equation}

The value of polarizability at time step n+1, $P_{NLs}^{n+1}$, is calculated as follow: First, the wave vector $k^{n+1}$ is calculated by solving the differential equation (\ref{s4}), in Yee FDTD algorithm. Second, the group velocity, $v^{n+1}$, corresponding to the value of the wavevector, $k^{n+1}$, is determined from the realistic conduction band structure of BP through the equation (\ref{s3}). Having the value of the $v^{n+1}$, the polarizability in the time step $n+1$ can be obtained by solving the differential equation of (\ref{s2}). For calculating the values in time step $n+1$ we need to use the stored values of the two time-steps earlier, $n-1$, fulfilling the central difference nature of the Yee FDTD algorithm where the finite-difference equations are central about the time point $t^n$. Electric field in the free space in the time step $n+1$ can be calculated by $E_s^{n+1}=D_s^{n+1}/\epsilon_0$. 

\begin{table}[h!]
\begin{center}
\caption{Parameters used in the simulation}
\label{table1}
\begin{tabular}{c c}
\hline
\textbf {$\omega_p$} & {660 cm$^{-1}\cong 19.79$ THz}  \\
%\textbf {} & {$\sim 806$ cm$^{-1}\cong 24.16$ THz at $T=10$ K} \\
\hline
\textbf {$\Gamma$} & {72.8 cm$^{-1}\cong 2.18$ THz} \\
\hline
{$m^*$ ($m_0=9.1$x10$^{-31}$ kg)} & {0.076 $m_0$ \cite{gaddemane18}} \\
\hline
\textbf {$\epsilon_{\infty}$} & {9.7} \\
\hline
\end{tabular}
\end{center}
\end{table}

In the simulation, the values for $\Gamma$ and $\omega_p$ were extracted from the experimental data, and  $\epsilon_\infty$ and $m^*$ are given from literature. The time step $\Delta t$ was set to 0.5 fs and the space step $\Delta z$ was set to 0.3 $\mu$m. All the parameters used in the FDTD simulations are listed in Table \ref{table1}.

According to the model, if the electron moves in a perfectly parabolic potential the nonlinear polarization component $P_{NL}$ vanishes according to eqs. (\ref{s2}) and (\ref{s3}). On the other hand, when the THz fields are intense enough as to drive electrons in anharmonic regions of the band dispersion nonlinear effects can not be neglected any more. 

\begin{figure*}
\begin{center}
\leavevmode
\includegraphics [width=15cm]{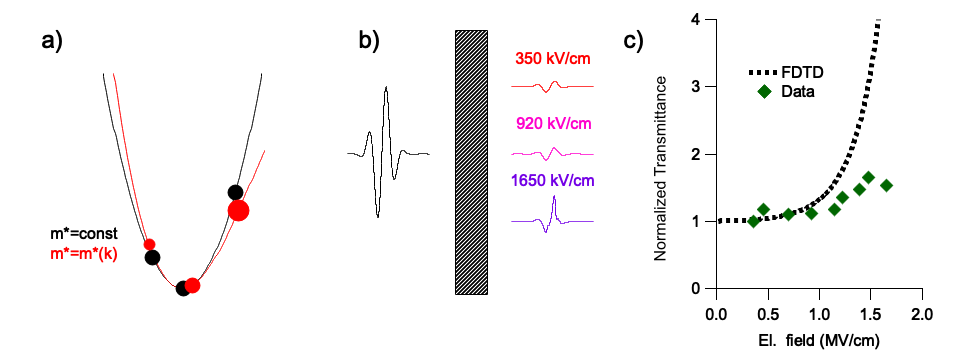}  
\end{center}
\caption{a) Schematics of the wavevector dependent mass variation. The effective mass being defined by $m^*=\frac{\hbar^2}{d^2\varepsilon/dk^2}=m^*(k)$, in a non parabolic potential as the one depicted here in red depends on the wavevector, while $m^*$ is a constant for parabolic bands.  b) Incoming and transmitted THz fields from the BP sample as from the FDTD calculation. The outcoming THz fields are normalized to the incoming peak value. With increasing incoming THz fields, the transmitted pulses also increase thereby qualitatively reproducing the experimentally observed saturable absorption. c) THz field dependence of the transmittance (measured at peak position) defined as $E_{out}^{max}/E_{in}^{max}$, and normalized to the lowest field value, for both FDTD and experiment.}
\label{Fig2}
\end{figure*}

This scenario can also be understood in terms of a wavevector-dependent effective mass. While for a parabolic potential $m^*=\hbar^2(\frac{\partial^2\varepsilon}{\partial k^2})^{-1}$ is a constant, in the case of anharmonicity $m^*=m^*(k)$ (see Fig. \ref{Fig2}a). In most realistic cases the electrons then start exploring regions of the Brillouin Zone, where the band dispersion flattens with respect to the band bottom, thereby resulting in an increased effective mass $m^*=m^*(k)$, and a saturable absorption behavior takes place. In the case of BP, the nonlinear FDTD calculation qualitatively reproduces the enhancement of transmitted THz pulses for increasing incoming fields. However, for fields higher than 1 MV/cm, the FDTD simulation strongly overestimates the expected transmitted intensity reduction (see Fig. \ref{Fig2}b and c).

\section{Two-temperature model calculation}
\label{2T}

\begin{figure*}
\begin{center}
\leavevmode
\includegraphics [width=9cm]{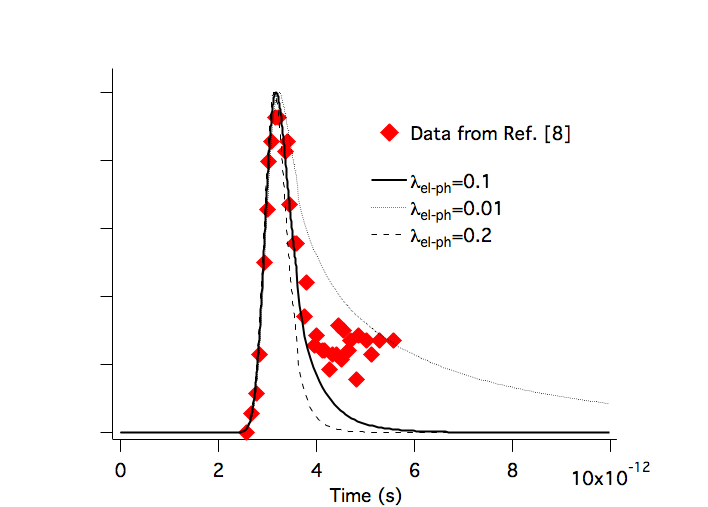}  
\end{center}
\caption{Time-resolved reflectivity data from Ref. \cite{montanaro22}, and corresponding simulations performed with the NTM.py code, by using the same material parameters as discussed above, and for different values of the electron-phonon coupling constant $\lambda_{el-ph}$. It is clear that $\lambda_{el-ph}\sim 0.1$ properly reproduces the experimental findings, and notably the short relaxation within $\sim 1$ ps.
}
\label{Tresolved}
\end{figure*}

\begin{figure*}
\begin{center}
\leavevmode
\includegraphics [width=9cm]{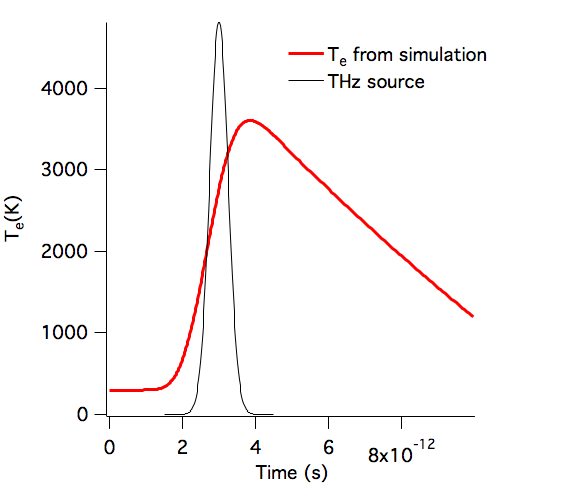}  
\end{center}
\caption{Time-dependent electronic temperature $T_e(t)$, calculated at the highest fluence for a Gaussian THz source centered at 3 ps (see text).  
}
\label{Tsim}
\end{figure*}

The calculation of the electron temperature $T_e$ reached after THz photoexcitation has been performed with the NTMpy package \cite{alber20}, based on a two-temperature model. The model assumes that two independent reservoirs ($T_e$ for electrons, and $T_l$ for the lattice) are present, and exchange heat after absorption of the THz pulse. This takes place through the coupled differential equations:

\begin{equation}
c_e\rho^{BP}\frac{\partial T_e}{\partial t}=G (T_l-T_e)+S(t),
\end{equation}
\begin{equation}
c_l\rho^{BP}\frac{\partial T_l}{\partial t}=G (T_e-T_l), 
\end{equation}

where $z$ indexes the stacked BP layers and:
\begin{itemize}
\item{$c_e=\gamma_eT_e/\rho^{BP}$ is the electron heat capacity, with $\gamma_e=\frac{1}{3}\pi^2D(\varepsilon_F)k_B^2$,}
\item{$c_l=9nk_B(\frac{T}{\theta_{Debye}})^3\int_0^{\theta_{Debye}/T}\frac{x^4e^xdx}{(e^x-1)^2}$ is the lattice heat capacity, }
\item{$G=3\gamma_e\frac{\lambda_{e-ph}E_{Debye}^2}{\pi\hbar k_B}$ is the electron phonon coupling constant,}
\item{$S(t)$ corresponds to the heat provided by the THz pulse, which is considered here as a Gaussian,}
\end{itemize}
with $\rho^{BP}$= 2600 kg/m$^3$, $\theta_{Debye}$=267 K ($E_{Debye}$=0.023 eV). $n=5.3\times10^{28}$ m$^{-3}$ is the density of ions in the crystal. $\lambda_{el-ph}$ is the electron phonon coupling constant. 

The code calculates the time-evolution of $T_e$ and $T_l$ as a function of time, and for the various BP layers along the full 30  $\mu$m-thick stack, by taking into account the absorption of the THz beam at the various layers. A weighted average is finally performed to calculate an effective $T_e(t)=\frac{\Sigma_z a_z T_e^z(t)}{\Sigma_z a_z}$ by multiplying the electronic temperature of each layer by a coefficient $a_z=exp(-z/z_0)$, where $z_0$ is the penetration depth at 1 THz.

As thoroughly discussed in \cite{allen},  the most reliable experimental technique to evaluate the electron-phonon coupling constant $\lambda_{el-ph}$ is through {\it pump-probe} experiments. In order to provide an estimate for  $\lambda_{el-ph}$, we first run the NTM.py code on the time-resolved measurement from Ref. \cite{montanaro22}. To this aim we employ the same material's parameters as discussed above while trying different values for  $\lambda_{el-ph}$. It turns out that $\lambda_{el-ph}\sim0.1$ provides a relaxation time in quite good agreement with respect to the time-resolved reflectivity data. On the other hand, for $\lambda_{el-ph}^{min}=0.01$ or $\lambda_{el-ph}^{max}=0.2$ the results provided by the NTM.py code start showing some deviations in the relaxation behavior with respect to the experiment.

Once the electron-phonon coupling constant $\lambda_{el-ph}$ has been established we can finally apply the code to estimate the electronic temperature $T_e$ in the present experiment. To this aim we simulate our source with a Gaussian with $\Delta t^{FWHM}=$ 590 fs. The fluence varies from 1.4 to 30.8 J/m$^2$. We take as a central frequency $\nu_0$ of our source $S(\nu)$ the first moment
\begin{equation}
\nu_0=\frac{\int \nu S(\nu)d\nu}{\int S(\nu)d\nu}=1.13 \: {\rm THz}.
\end{equation}
We finally obtain $\lambda_0=c/\nu_0=295$ $\mu$m. With these parameters we calculate $T_e(t)$, as reported in Fig. \ref{Tsim}. In our experiment we address the non-linear electrodynamic properties as a function of the THz fluence. This implies that we are not interested in $T_e(t)$ at times retarded with respect to the THz pulse S(t). What matters for the interpretation of our results is the $T_e$ value which is being probed by the THz pulse itself. To this aim we calculate an effective electronic temperature
\begin{equation}
T_e=\frac{\int T_e(t)S(t)dt}{\int S(t)dt},
\end{equation}
at each fluence and for both orientations. These are the temperature values employed in Fig. 2 of the main text.

\section{Spectral Weight's temperature dependence}
\label{SW_T}

Because of charge conservation, the chemical potential can be defined as the value satisfying the normalization condition \cite{ashcroft}:

\begin{equation}
N=\int\frac{1}{4\pi^3}f(\varepsilon(\mathbf{k}))\rm{d}\mathbf{k},
\end{equation}
where 
\begin{equation}
f(\epsilon(\mathbf{k}))=\frac{1}{e^{[\varepsilon(\mathbf{k})-\mu(T)]/k_BT}+1}.
\end{equation}

\begin{figure*}
\begin{center}
\leavevmode
\includegraphics [width=15cm]{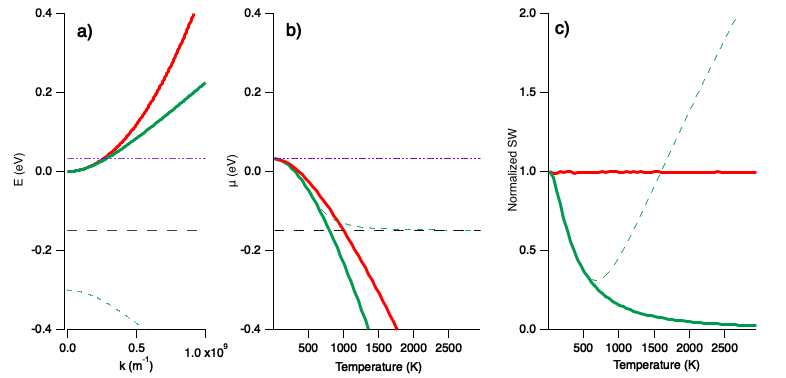}  
\end{center}
\caption{a) Energy dispersion for three different energy band dispersions, and correspondingly calculated b) chemical potential $\mu(T)$, and c) spectral weight $SW(T)$. The red lines correspond to parabolic dispersion, while green lines correspond to massive Dirac dispersion when the conduction band only is taken into account. The dashed green lines correspond to massive Dirac dispersion for symmetric conduction and valence bands with a 0.3 eV gap, thus allowing the formation of electron-hole pairs. The purple dashed-dotted line indicate the Fermi level, while the dashed black line shows the center of the gap.
}
\label{mu_T}
\end{figure*}

Once $N$ is known, it is possible to numerically evaluate the full temperature dependence of $\mu(T)$ for any possible energy band dispersion  $\varepsilon=\varepsilon(\mathbf{k})$. We report in Fig. \ref{mu_T}b, the temperature dependence of $\mu(T)$, for two different band dispersions: One parabolic band dispersion (red)  $\varepsilon(k)=\frac{\hbar^2k^2}{2m}$, with $m=0.08\cdot m_e$, and a relativistic Dirac band dispersion  (green) in the form $\varepsilon(k)=\sqrt{m^2c^4+\hbar^2c^2k^2}$, with $m=0.08\cdot m_e$, and $c= 4.7\times10^5$ m/s, as fits to the black phosphorus band structure along the {\it armchair} direction. For both dispersions the chemical potential rapidly decreases with temperature, and crosses the conduction band bottom already at about 350 K and 300 K for parabolic and Dirac dispersions respectively. While the trend of the chemical potential is quite similar for the two dispersions, the difference in the behavior of the $SW$ as seen in  (Fig. \ref{mu_T}c) is stunning. For the parabolic potential the $SW$ remains constant at all temperatures, while it dramatically decreases and saturates close to 0 for the Dirac dispersion.

Interestingly, if we include the presence of a valence band symmetric with respect to the conduction band the situation drastically changes both from the point of view of the chemical potential and from that of the $SW$. We performed the calculation (green dashed line) in the case of the Dirac dispersion, by assuming a gap of 0.3 eV (independent on the value of $m$). In this case $\mu(T)$, first drops similarly to what observed for the single band calculation, and then saturates for an energy of  -$0.15$ eV, corresponding to the center of the gap. From the point of view of the $SW$, in correspondence with the saturation, we observe an upturn of $SW(T)$, which asymptotically increases linearly with $T$, as a consequence of the $T$-induced formation of electron-hole pairs.

\end{document}